\documentclass[11pt]{article}
\usepackage{epsfig,amsmath,amssymb,a4}

\begin{document}

\title{\bf $\pi\eta$ scattering in generalized chiral perturbation theory\setcounter{footnote}{4}\thanks{Presented by J.
N. at Int. Conf. Hadron Structure '02, Herlany, Slovakia,
September 22-27, 2002 }}
\author{J. Novotn\'{y}\setcounter{footnote}{1}\footnote{Jiri.Novotny@mff.cuni.cz}
and M. Koles\'{a}r\footnote{Marian.Kolesar@mff.cuni.cz} \\
Institute of Particle and Nuclear Physics,\\Charles University, V
Hole\v{s}ovi\v{c}k\'{a}ch 2, 180 00 Czech Republic}
\date{}\maketitle
%
\begin{abstract}{We calculate the amplitude of $\pi\eta$ scattering
in generalized chiral perturbation theory at the order $O(p^4)$
and present a preliminary results for the numerical analysis of
the S-wave scattering length, which seems to be particularly
sensitive to the deviations from the standard case.}
\end{abstract}
\section{Generalized Chiral Perturbation Theory}

%
%
%

As it is well known, on the classical level the QCD Lagrangian with $N_f$ {%
massless} quarks (corresponding to so-called chiral limit of QCD) is
invariant w.r.t. {chiral symmetry }($\chi S)${\ group }$SU(N_f)_L\times
SU(N_f)_R$. On the quantum level there exist strong theoretical (for $%
N_f\geq 3$) and phenomenological arguments for spontaneous symmetry
breakdown (SSB) of $\chi S$ according to the pattern $SU(N_f)_L\times
SU(N_f)_R\rightarrow SU(N_f)_V$. As a consequence of Goldstone theorem, $%
N_f^2-1$ pseudoscalar Goldstone bosons (GB) appear in the particle spectrum
of the theory. These massless pseudoscalars dominate the low energy dynamics
of QCD and interact weekly at low energies $E<<\Lambda _H$, where $\Lambda
_H\sim 1GeV$ is the hadronic scale corresponding to the masses of the
lightest nongoldstone hadrons. The most important order parameters of this
pattern of SSB are the Goldstone boson decay constant $F_0$ and the quark
condensate\footnote{%
The parameter $F_0$ is however more fundamental in the sense that $F_0\neq 0$
means both necessary and sufficient condition for SSB, while $\langle
\overline{q_f}q_f\rangle _0\neq 0$ corresponds to the sufficient condition
only.} $\langle \overline{q_f}q_f\rangle _0$.

Within the real QCD the quark mass term ${\cal L}_{f,mass}^{QCD}\,$breaks $%
\chi S$ explicitly. The GB become pseudogoldstone bosons (PGB) with nonzero
masses. Nevertheless, for $m_f<<\Lambda _H$, ${\cal L}_{f,mass}^{QCD}$ can
be treated as a small perturbation. As a consequence, the PGB masses $M_P$
can be expanded in the powers (and logarithms) of the quark masses and the
interaction of PGB at energy scale $E<<\Lambda _H$ continues to be weak. PGB
are identified with $\pi ^0,\pi ^{\pm }\,$ for $N_f=2$ and $\pi ^0,\pi ^{\pm
},K^0,\overline{K^0},K^{\pm },\eta $ for $N_f=3$. Because $M_P<\Lambda _H$,
the QCD dynamics at $E<<\Lambda _H$ is still dominated by these particles
and can be described in terms of an effective theory known as chiral
perturbation theory ($\chi PT$). The Lagrangian of $\chi PT$ can be
constructed on the base of symmetry arguments only; the unknown information
about the nonperturbative properties of QCD are hidden in the parameters
known as low energy constants (LEC)\cite{CHPT1}.

In order to be able to treat the effective theory as an expansion in powers
of $(p/\Lambda _H)$ (where $p$ are generic external momenta) and $%
(m_f/\Lambda _H)$, it is necessary to assign to each term ${\cal L}%
^{(m,n)}=O(\partial ^mm_f^n)$ of the effective Lagrangian a single parameter
called {\em chiral order}. To the terms ${\cal L}_k$ with chiral order $k$
it is referred as to $O(p^k)$ terms. Obviously, $\partial =O(p).$ The matter
of discussion is, however, the question concerning the chiral power of $m_f$%
. This question is intimately connected to the scenario according which the
SSB of $\chi S$ is realized.

The standard scenario corresponds to the assumption, that the SSB order
parameter $\langle \overline{q}q\rangle _0$ is large in the sense, that the
ratio $X$%
\begin{equation}
X=\frac{2B_0\widehat{m}}{M_\pi ^2}  \label{X}
\end{equation}
(where $B_0=-\langle \overline{q}q\rangle _0/F_0^2$ and $\widehat{m}%
=(m_u+m_d)/2$) is close to one. Because $M_\pi ^2=O(p^2)$, it is then
natural to take $m_f=O(p^2)$, i.e. $k=m+2n$. This results into the Standard $%
\chi PT$ ($S\chi PT$ in what follows)\cite{CHPT2}. This scenario has been
experimentally confirmed \cite{E865} for $N_f=2$ and it is perfectly
compatible with experiment for $N_f=3$ in $\pi ,\,K$ sector.

Let us note that at $O(p^2)$ there is none free parameter, because at this
chiral order $F_0=F_\pi =93.2{\rm MeV}$, $\,\,2B_0\widehat{m}=M_\pi ^2=135%
{\rm MeV}$ and $r=m_s/\widehat{m}\simeq 26$.

Alternative to this way of chiral power counting is Generalized
$\chi PT$ ($G\chi PT$) \cite{CHPT3} corresponding to the scenario
with
small quark condensate $X<<1$ so that it is natural to take $m_f=O(p)$ and $%
B_0=O(p)$. {\em I.e.} $k=m+n$ and the $O(p^2)$ Lagrangian is
\begin{eqnarray}
{\cal L}_2 &=&\frac{F_0^2}4\left( \langle \partial _\mu U^{+}\partial ^\mu
U\rangle +2B_0\langle U^{+}{\cal M}+{\cal M}^{+}U\rangle +A_0\langle (U^{+}%
{\cal M})^2+({\cal M}^{+}U)^2\rangle \right.  \nonumber \\
&&+\left. Z_0^P\langle U^{+}{\cal M}-{\cal M}^{+}U\rangle ^2+Z_0^S\langle
U^{+}{\cal M}+{\cal M}^{+}U\rangle ^2\right) .  \label{L2GCHPT}
\end{eqnarray}
This scenario is still possible for $N_f=3$, as has been discussed in \cite
{Descotes}\footnote{%
The point is, that provided we define the $n$-flavor condensate as
\[
\langle \overline{q}q\rangle _0^{(n)}=\lim_{m_f\rightarrow 0,\,f\leq
n}\langle \overline{q}q\rangle _0,
\]
the two-flavor condensate relevant for the $\chi PT$ with $N_f=2$ is related
to the three-flavor one relevant for the $\chi PT$ with $N_f=3$,
\[
\langle \overline{u}u\rangle _0^{(2)}=\langle \overline{u}u\rangle
_0^{(3)}-m_s\overline{Z}_1^S+\ldots
\]
where $\overline{Z}_1^S$ is the fluctuation parameter measuring a violation
of the Zweig rule in the $0^{++}$ channel. That means, that the three-flavor
condensate might be small provided $\overline{Z}_1^S$ is large. Recent
phenomenological studies suggest possibility of $X^{(3)}\sim 1/2$, cf. \cite
{Descotes} and \cite{Moussallam}.}.

In the generalized case, there are two free parameters in the $O(p^2)$
effective Lagrangian, the usual choice is $(r,\zeta =Z_0^S/A_0)$. Within $%
S\chi PT$ at $O(p^2)$ we get roughly $(r,\zeta)=(26,0)$. Note, that $\zeta $
measures the violation of Zweig rule in the $0^{++}$ channel, which is,
however, not well under control. In what follows, we therefore prefer to
parametrize the deviations from $S\chi PT$ directly on terms of $X$,{\em \
i.e.} our choice of free parameters is $(r=m_s/\widehat{m},\,\,X)$, cf. (\ref
{X}) (standard $O(p^2)$ values are then $(26,1)$). The $O(p^2)$ LEC can be
then expressed in terms of these free parameters.

In oder to distinguish between the two scenarios of $\chi S$ SSB,
it is necessary to find observables, which are sensitive to the
deviations from the standard case. It seems, that the $\pi \eta $
scattering might offer such observables, though we left open the
question about their experimental accessibility.

Let us note, that the amplitude of this process was already calculated
within $S\chi PT$ to $O(p^4)$ (and within the extended $S\chi PT$ with
explicit resonance fields) in the paper \cite{etapi}, where the authors
presented prediction for the scattering lengths and phase shifts of the $S$,
$P$ and $D$ partial waves. We quote here their $O(p^4)$ results for the $S$%
-wave scattering length $a_0$ (in the units of the pion Compton wavelength):
$a_0^{S\chi PT}=7.3\times 10^{-3}$ and $a_0^{S\chi PT+resonances}=4.9\times
10^{-3}$.

\section{{\bf General structure of the }$\pi \eta ${\bf \ scattering
amplitude}}

Due to isospin conservation and Bose symmetry, the process is described in
terms of one $s-u$ symmetric invariant amplitude $A(s,t;u)$
\[
\langle \pi ^b\eta _{out}|\pi ^a\eta _{in}\rangle ={\rm i}(2\pi )^4\delta
(P_f-P_i)\delta ^{ab}A(s,t;u)
\]
Using analyticity, unitarity, crossing symmetry and assuming chiral
expansion in the same way as in \cite{pipi} we get the following general
form of the $O(p^4)$ amplitude\footnote{%
Here and in what follows, $\Delta =M_\eta ^2-M_\pi ^2$ and $\Sigma =M_\eta
^2+M_\pi ^2$ .}
\begin{eqnarray}
A(s,t;u) &=&R^{(3)}(t;s,u)+V_0(t)+W_0(s)+W_0(u)  \nonumber \\
&&+[(t-u)s+\Delta ^2]W_1(s)+[(t-s)u+\Delta ^2]W_1(u)  \label{general}
\end{eqnarray}
Here $R^{(3)}(t;s,u)$ is the most general $s-u$ symmetric subtraction
polynomial of the third order\footnote{%
Within the generalized chiral expansion, $\alpha _{\pi \eta }=O(1)$, $\beta
_{\pi \eta }=O(p)$, $\lambda _{\pi \eta },\widetilde{\lambda }_{\pi \eta
}=O(p^2)$, $V_0,W_0,W_1$ $=$ $O(p^4)$ and $\kappa _{\pi \eta }$, $\widetilde{%
\kappa }_{\pi \eta }=$ $O(p^4)$.}
\begin{eqnarray}
R^{(3)}(t;s,u) &=&\frac 1{3F_\pi ^2}(\alpha _{\pi \eta }M_\pi ^2+\beta _{\pi
\eta }(t-\frac 23\Sigma )+\frac{\lambda _{\pi \eta }}{F_\pi ^2}(t-\frac
23\Sigma )+\frac{\widetilde{\lambda }_{\pi \eta }}{F_\pi ^2}(s-u)^2
\nonumber \\
&&+\frac{\kappa _{\pi \eta }}{F_\pi ^4}(s-u)^2(t-\frac 23\Sigma )+\frac{%
\widetilde{\kappa }_{\pi \eta }}{F_\pi ^4}(t-\frac 23\Sigma )^3).  \label{R}
\end{eqnarray}
The unitarity corrections $V_0$,$W_0$, $W_1$ start at $O(p^4)$ and are
determined by means of the dispersion integrals along the cuts $((M_\pi
+M_\eta )^2,\infty )$ or $(4M_\pi ^2,\infty )$ with the discontinuities
given by the right hand cut discontinuities of the $S$ and $P$ partial waves
in the $s$ and $t$ channel. Using partial waves unitarity, it is possible to
proceed iteratively and determine the relevant $O(p^4)$ discontinuities
through the $O(p^2)$ amplitudes $A^{\eta \pi \rightarrow \phi _\alpha \phi
_\beta }$, $A^{\phi _\alpha \phi _\beta \rightarrow \eta \pi }$, $A^{\pi \pi
\rightarrow \phi _\alpha \phi _\beta }$ and $A^{\eta \eta \rightarrow \phi
_\alpha \phi _\beta }$. These are real and first order polynomials in $%
s,\,t,\,u$, so that we can parametrize them with (altogether 11) real
parameters. Adding to this the 2 extra real coefficients of $O(p^4)$ part of
the subtraction polynomial $R^{(3)}(t;s,u)$, it is possible to parametrize
the $O(p^4)$ amplitude $A(s,t;u)$ in terms of 13 real free parameters%
\footnote{%
In the following formulas, the parameters of the $O(p^2)$ amplitudes $%
A^{\eta \pi \rightarrow \phi _\alpha \phi _\beta }$ and $A^{\phi _\alpha
\phi _\beta \rightarrow \eta \pi }$ are $\alpha _{\pi \eta }$, $\beta _{\pi
\eta }$, $\alpha _{\pi \eta K}$, $\beta _{\pi \eta K}$ for
\par
$\phi _\alpha \phi _\beta =\pi \eta ,\,\overline{K}K$, the parameters of $%
O(p^2)$ amplitudes $A^{\pi \pi \rightarrow \phi _\alpha \phi _\beta }$ and $%
A^{\eta \eta \rightarrow \phi _\alpha \phi _\beta }$ are $\alpha _{\pi \pi }$%
, $\beta _{\pi \pi }$, $\alpha _{\eta \eta }$, $\alpha _{\pi K}$, $\beta
_{\pi K}$, $\alpha _{\eta K}$, $\beta _{\eta K}$ for $\phi _\alpha \phi
_\beta =\pi \pi ,\eta \eta ,\,\,\overline{K}K$.}. As a result of the
iterative procedure we get for the $O(p^4)$ unitarity corrections $V_0$, $W_0$%
, $W_1$ the following formulas: $W_1^{(4)}(s)=0$ and {\small {\
\begin{eqnarray*}
W_0^{(4)}(s) &=&\overline{J}_{\pi \eta }(s)\left( \frac 1{3F_\pi ^2}\alpha
_{\pi \eta }M_\pi ^2\right) ^2 \\
&&+\overline{J}_{KK}(s)\frac 3{8F_\pi ^4}[\beta _{\pi \eta K}(s-\frac
13\Sigma -\frac 23M_K^2)-\frac 13(2M_K^2-\Sigma +\alpha _{\pi \eta K}M_\pi
^2)]^2 \\
V_0^{(4)}(s) &=&\overline{J}_{\pi \pi }(s)\frac 1{3F_\pi ^4}\alpha _{\pi
\eta }M_\pi ^2[\beta _{\pi \pi }(s-\frac 43M_\pi ^2)+\frac 56\alpha _{\pi
\pi }M_\pi ^2] \\
&&-\overline{J}_{\eta \eta }(s)\frac 1{18F_\pi ^4}\alpha _{\eta \eta }\alpha
_{\pi \eta }M_\pi ^4\left( 1-\frac{4M_\eta ^2}{M_\pi ^2}\right) \\
&&+\overline{J}_{KK}(s)\frac 1{8F_\pi ^2}[\beta _{\pi K}(s-\frac 23M_\pi
^2-\frac 23M_K^2)+\frac 23((M_K-M_\pi )^2+2\alpha _{\pi K}M_KM_\pi )] \\
&&\times [\beta _{\eta K}(3s-2M_K^2-2M_\eta ^2)+\alpha _{\eta K}(2M_\eta
^2-\frac 23M_K^2)]
\end{eqnarray*}
}}where $\overline{J}_{PQ}(s)$ is the Chew-Mandelstam function, cf.\cite
{pipi}. The role of $\chi PT$ is then reduced to the determination of the
above mentioned parameters in terms of LEC and quark masses. Let us now
briefly comment on the results of the calculations.

\section{$\pi \eta ${\bf \ amplitude in }$G\chi PT$ at $O(p^4)$}

Let us write the complete $O(p^4)$ amplitude $A(s,t,u)$ in the form
\[
A(s,t,u)=A^{(2)}(s,t,u)+A^{(3)}(s,t,u)+A^{(4)}(s,t,u),\,\,\,\,\,{\rm where}%
\,\,\,A^{(k)}(s,t,u)=O(p^k).
\]
$A^{(2)}$ and $A^{(3)}$ contain the contributions from tree graphs with
vertices derived form Lagrangians ${\cal L}_2$ and ${\cal L}_3=O(p^3)$
respectively (see (\ref{L2GCHPT}) and {\em e.g.} \cite{GCHPT}); $A^{(4)}$
includes tree graphs with vertices from ${\cal L}_4=O(p^4)$ as well as
1-loop graphs with vertices from ${\cal L}_2$.

Because the unitarity corrections start at $O(p^4)$, $A^{(2)}(s,t,u)$ and $%
A^{(3)}(s,t,u)$ are both polynomials of the form (cf. (\ref{R}))
\[
A^{(k)}(s,t,u)=\frac 1{3F_\pi ^2}(\alpha _{\pi \eta }^{(k)}M_\pi ^2+\beta
_{\pi \eta }^{(k)}(t-\frac 23(M_\pi ^2+M_\eta ^2)),\,\,\,\,\,k=1,2.
\]
Moreover, the amplitude must vanish in the chiral limit, therefore $\beta
_{\pi \eta }^{(2)}=0$. From (\ref{L2GCHPT}) we get
\[
A^{(2)}(s,t,u)=\frac{M_\pi ^2}{3F_\pi ^2}\alpha _{\pi \eta }^{(2)},
\]
where, in terms of the free parameters $(r,X)$

\[
\alpha _{\pi \eta }^{(2)}=1+\frac{\,\left( 1+2\,r\right) \,\left( 2\,\left(
1-X\right) +r\varepsilon (r)\right) }{\,\left( 2+r\right) }-\frac{2\Delta
_{GMO}}{r-1}.
\]
In this formula
\begin{equation}
\varepsilon (r)=2\frac{r_2-r}{r^2-1},\,\,\,r_2=2\frac{M_K^2}{M_\pi ^2}%
-1,\,\,\,\Delta _{GMO}=\frac{3M_\eta ^2+M_\pi ^2-4M_K^2}{M_\pi ^2}\simeq
-3.6.
\end{equation}
The standard $O(p^2)$ result (corresponding to the current algebra
(CA)) corresponds to $\alpha _{\pi \eta }^{(2)}=1$ \cite{etapi}.
The dependence of $\alpha _{\pi \eta }^{(2)}$ on $r$ and $X$ is
shown in Fig. 1. The deviation from the standard case might be
even by a factor ten larger than the standard value, provided the
quark mass ratio $r$ is small in comparison with $r_2$ and the
three-flavor ratio $X$ is smaller than one. This result is
encouraging enough to calculate the higher order corrections.

\begin{figure}[tbp]
\hspace{1.5cm}\epsfig{figure=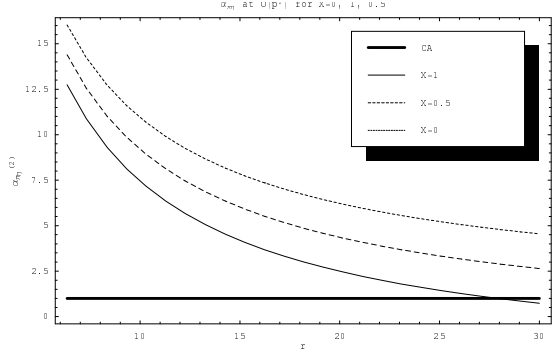,width=0.6\textwidth}
\caption{The dependence of the parameter $\alpha _{\pi \eta }^{(2)}$ on $r$
for $X=1,\,\,0.5$ and $0$. The thick solid line corresponds to current
algebra result.}
\label{My label}
\end{figure}

The $O(p^3)$ Lagrangian ${\cal L}_3$ contains 9 LEC\footnote{%
Some of them, namely $\widetilde{\xi },\,\rho _3$,\,$\rho
_4,\,\rho _5,\,\rho _6,\,\rho _7$, violate Zweig rule and might be
neglected (the only exception is $\widetilde{\xi }$, which could
be relevant as the measure of the fluctuations in the $0^{++}$
channel, cf. \cite{Descotes}).}, namely $\xi $, $\widetilde{\xi }$
and $\rho _i$, $i=1,\ldots ,7$. $\xi $ can be expressed in terms
of decay constants as follows
\[
\widehat{m}\xi =\frac{\Delta _F}{r-1},\,\,\,\,\,\,\,\Delta _F=\frac{F_K^2}{%
F_\pi ^2}-1\simeq 0.5.
\]
For the NLO corrections in terms of $X$, $\varepsilon $, $\Delta _{GMO}$, $%
\Delta _F$, $\widetilde{\xi }$ and remaining $O(p^3)$ LEC we get \
\
\begin{eqnarray*}
\beta _{\pi \eta }^{(3)} &=&4\widehat{m}\widetilde{\xi }(r+2) \\
\alpha _{\pi \eta }^{(3)} &=&\frac{2\Delta _F}{(r-1)}(1-\frac{2(2r+1)^2(1-X)%
}{3(r+2)}-\frac{(8+r(6r+13))(r-1)\varepsilon }{6(r+2)}-\frac{\Delta _{GMO}}3)
\\
&&-4\widehat{m}\widetilde{\xi }(2r+1)\left( 1-(X-1)+\frac{r^2-1}{2r+1}%
\varepsilon +\frac{\Delta _{GMO}}{2r+1}\right)\\&& +\frac 83\widehat{m}%
\widetilde{\xi }(r+2)\frac \Sigma {M_\pi ^2}+\ldots
\end{eqnarray*}
The ellipses stay for the $O(m_f^3)$ terms which includes the unknown LEC $%
\rho _i$.

\begin{figure}[tbp]
\vspace{0.1cm} \hspace{-0.7cm}\epsfig{figure=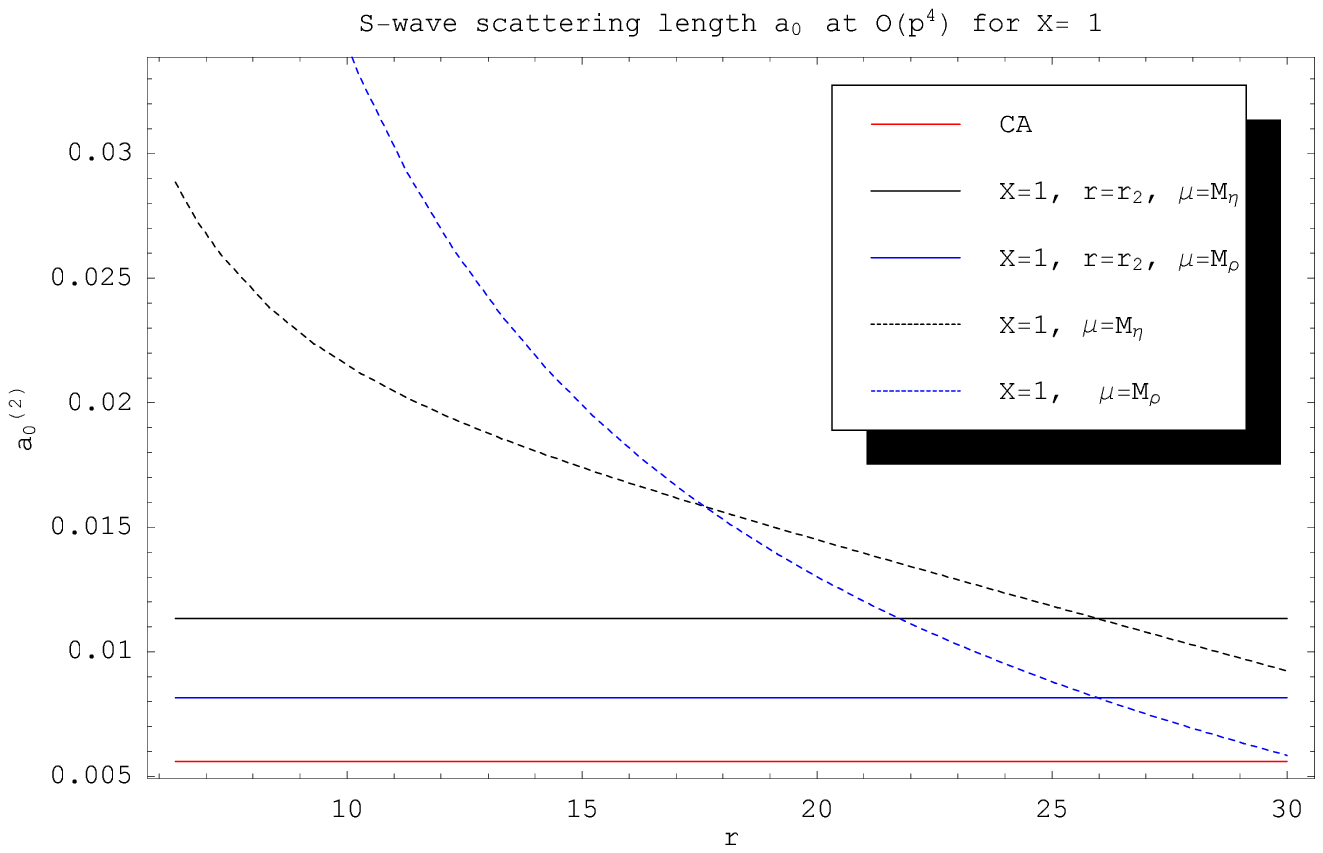,width=0.49\textwidth}
\hspace{0.8cm} \epsfig{figure=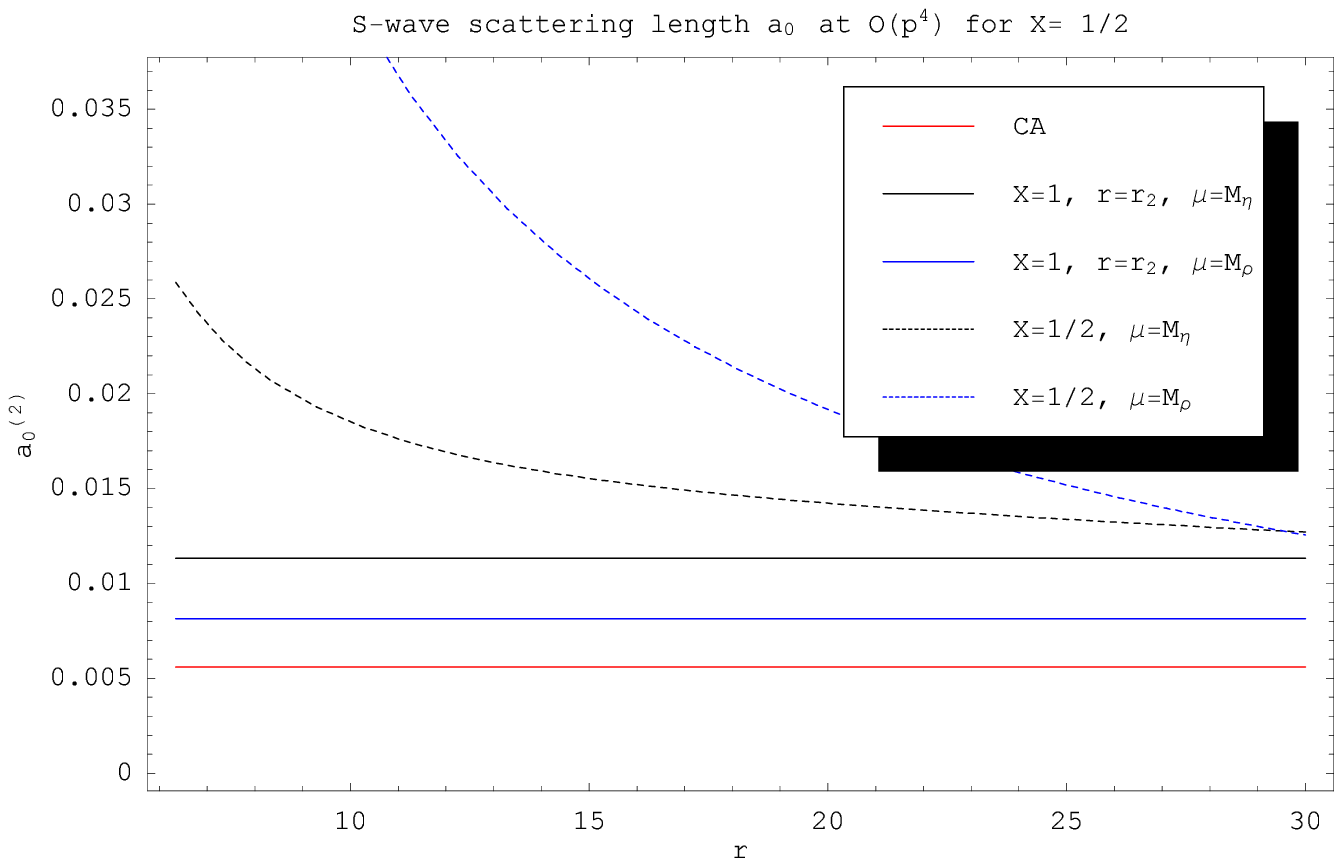,width=0.49\textwidth}
\caption{The dependence of the $S$-wave scattering length on $r$ for $X=1$
and $X=0.5$ with $\mu=M_\eta$ and $\mu=M_\rho$ . The solid lines mimic the $%
S\chi PT$ ($X=1$ and $r=r_2$) at $O(p^4)$. The lowest line corresponds to
current algebra result.}
\label{2}
\end{figure}

The NNLO $O(p^4)$ corrections have the general form (cf. (\ref{general}))
with $\kappa =\widetilde{\kappa }=0$ and {\small {\ }} with $V_0^{(4)}$ and $%
W_i^{(4)}$ given above. The complete formulas for $A^{(4)}$, which are
rather lengthy, will be published elsewhere. Let us only briefly comment on
the result.

As usual, 1-loop graphs which contribute to $A^{(4)}$ are generally
divergent; therefore the renormalization procedure is needed. As a result,
the amplitude depends explicitly on the renormalization scale $\mu $. This
scale dependence is compensated by the implicit scale dependence of the
renormalized LEC, this fact we used as a nontrivial check of our
calculations.

Let us also note, that the $O(p^4)$ Lagrangian is parametrized by means of
40 LEC, most of them are unknown. Influence of these unknown constant (as
well as that of the unknown $O(p^3)$ LEC) can be roughly estimated using the
above mentioned explicite $\mu $ dependence of the amplitude. The idea
behind is based on the assumption that the variation of LEC with $\mu $ is
of the same order as LEC themselves. Setting all the unknown LEC equal to
zero and varying the scale $\mu $ is then (up to the sign) equivalent to the
variation of the LEC and gives therefore information on the impact of the
unknown LEC.

The observable which seems to be sensitive to the deviation from the $S\chi
PT$ is the $S-$wave scattering length $a_0$, (note, that at $O(p^2)$, $%
a_0\propto \alpha _{\pi \eta }^{(2)}$, while $P-$wave scattering length $a_1$
starts at $O(p^3)$ ). In Fig. 2 we show the numerical results for $a_0 $ as
a function of the quark mass ratio $r$ for $X=1$ and $X=0.5\ $with $\mu
=M_\eta $ and $\mu =M_\rho $.


\section{Conclusions}

We have calculated the $\pi \eta $ scattering amplitude in $G\chi PT$ to the
order $O(p^4)$ and evaluated the $S-$wave scattering length as a function of
the free parameters $r$ and $X$. The influence of the other unknown LEC was
estimated using explicit dependence of the loops on the renormalization
scale $\mu $. Preliminary numerical results suggest that $S-$wave scattering
length might be sensitive to the values of the quark condensate and quark
mass ratio. $G\chi PT$ allows for systematically larger values of $a_0$ in
comparison to the standard case \cite{etapi}. The dependence of the loop
corrections on the renormalization scale {can be understood as a signal for
relatively strong dependence on the unknown LEC. }In order to get sharper
prediction further estimates will be necessary (resonances, sum rules...).
Numerical analysis of other observables, which has not been completed yet,
might be also interesting.

{\bf Acknowledgement. }This work was supported by the program ``Research
Centers'' (project number LN00A006) of the Ministry of Education of the
Czech Republic.

\end{document}